\documentclass[11pt,twoside]{article}
\usepackage{asp2010}

\aspvoltitle{9$^{th}$ Pacific Rim Conference on Stellar Astrophysics}
\aspvolauthor{Shengbang Qian, ed.}
\aspvolume{number not yet assigned}
\aspcpryear{2011}

\resetcounters

\begin{document}

\title{Red Dwarf Stars: Ages, Rotation, Magnetic Dynamo Activity and the
Habitability of Hosted Planets}
\author{Scott G. Engle$^1$ and Edward F. Guinan$^1$
\affil{$^1$Villanova University Dept. of Astronomy \& Astrophysics,
800 E. Lancaster Ave, Villanova, PA 19085, USA: sengle01@villanova.edu}}

\begin{abstract}
We report on our continued efforts to understand and delineate the magnetic
dynamo-induced behavior/variability of red dwarf (K5 V -- M6 V) stars over
their long lifetimes.  These properties include: rotation, light variations
(from star spots), coronal--chromospheric XUV activity and flares.  This
study is being carried out as part of the NSF-sponsored {\em Living with a Red
Dwarf} program\footnote{The Program website can be found at
\url{http://astronomy.villanova.edu/LWARD}}. The {\em Living with a
Red Dwarf} program's database of dM stars with photometrically determined
rotation rates (from starspot modulations) continues to expand, as does the
inventory of archival XUV observations.  Recently, the photometric properties 
of several hundred dM stars from the {\em Kepler} database are being 
analyzed to determine the rotation rates, starspot areal coverage/distributions
and stellar flare rates.  When all data setsare combined with ages from 
cluster/population memberships and kinematics, the determination of 
Age-Rotation-Activity relationships is possible. Such relationships have broad
impacts not only on the studies of magnetic dynamo theory and angular momentum loss 
of low-mass stars with deep convective zones, but also on the suitability of 
planets hosted by red dwarfs to support life.  With intrinsically low 
luminosities ($L < 0.02 L_\odot$), the liquid water habitable zones (HZs) 
for hosted planets are very close to their host stars -- typically at 
$\sim$0.1 AU $<$ HZ $<$ 0.4 AU. Planets located close to their host stars risk 
damage and atmospheric loss from coronal \& chromospheric XUV radiation,
 flares and plasma blasts via strong winds and coronal mass ejections.  In addition, 
our relationships permit the stellar ages to be
determined through measures of either the stars' rotation
periods (best way) or XUV activity levels. This also permits a determination of
the ages of their hosted planets.  We illustrate this with examples of age
determinations of the exoplanet systems: GJ  581 and HD 85512 (both with large
Earth-size planets within the host star's HZ), GJ 1214 (hot, close-in transiting 
super-Earth planet) and  HD 189733 (short period, hot-Jupiter planet interacting 
with its host star -- age from its dM4 star companion). 
\end{abstract}

\section{Background \& Introduction}
Studying red dwarf stars (K5 V -- M6 V -- ``K/M stars'' hereafter) is important because
they are the most populous stars in the Galaxy ($>$80\% of all stars -- see Fig. 1a) and could be
hosts to numerous planets, some of which may be suitable for life. Even though the
 vast majority of stars being searched for planets are brighter F, G and early K stars, 
thus far $>$ 30 exoplanets have been found orbiting red dwarfs, many of which are super-Earths
($\sim$2-10 M$_\oplus$) -- \citet{mayor2009,engle2009}. One of the primary motivations for the 
{\em Living with a Red Dwarf} program is to investigate whether these numerous, cool, low 
mass, low luminosity stars with extremely long (Fig. 1b) main-sequence lifetimes ($>$100 Gyr) 
are suitable for life on hosted planets in their close-in liquid water habitable zones 
(``HZs'' $<0.4$ AU).  We cannot realistically extrapolate our solar-type results to K/M 
stars since they are fundamentally different (mass/luminosity) and have deep 
convective zones (CZs -- fully convective atmospheres cooler than $\sim$dM3.5). Though
the specific origin of magnetic activity in red dwarf stars is still debated, they
are theorized to operate under the $\alpha^2$ (turbulent) dynamo, as opposed
to the $\alpha$-$\omega$ (shear interface) dynamo of the Sun and other
solar-type stars.  In theory, the $\alpha^2$ dynamo is driven purely by
convective motion, where the $\alpha$-$\omega$ dynamo relies on convection
and differential rotation. Previous X-ray studies show red dwarf stars have very
``efficient'' dynamos, with $L_{\rm X}$/$L_{\rm bol}$ values and chromospheric UV
surface fluxes up to $\sim$100$\times$ those of comparable age (or rotation) solar type stars. 

%\vspace{-5mm}
\begin{figure}
\begin{center}
%\vspace{-17mm}
\includegraphics[width=0.5\textwidth]{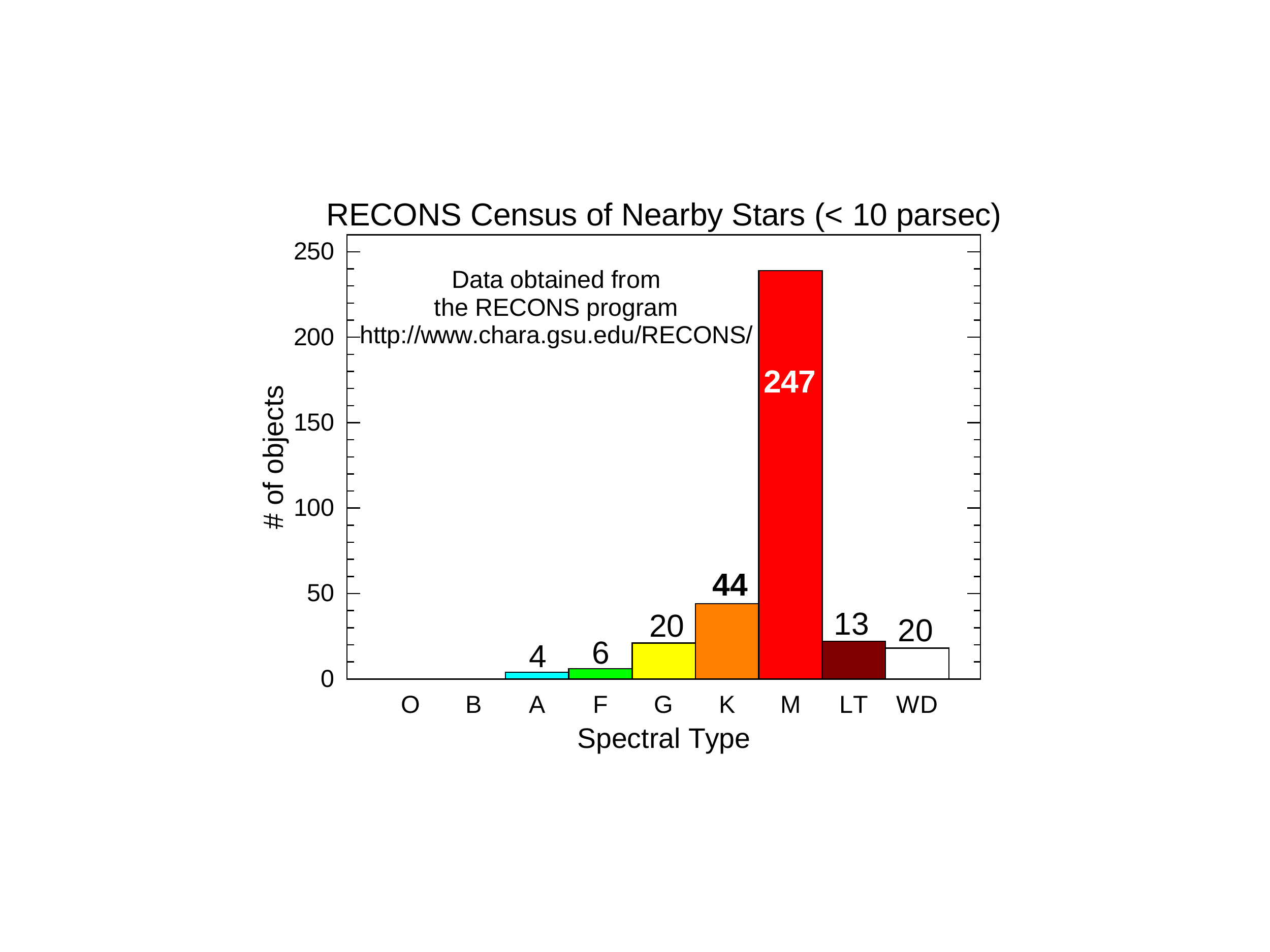}
\includegraphics[width=0.48\textwidth]{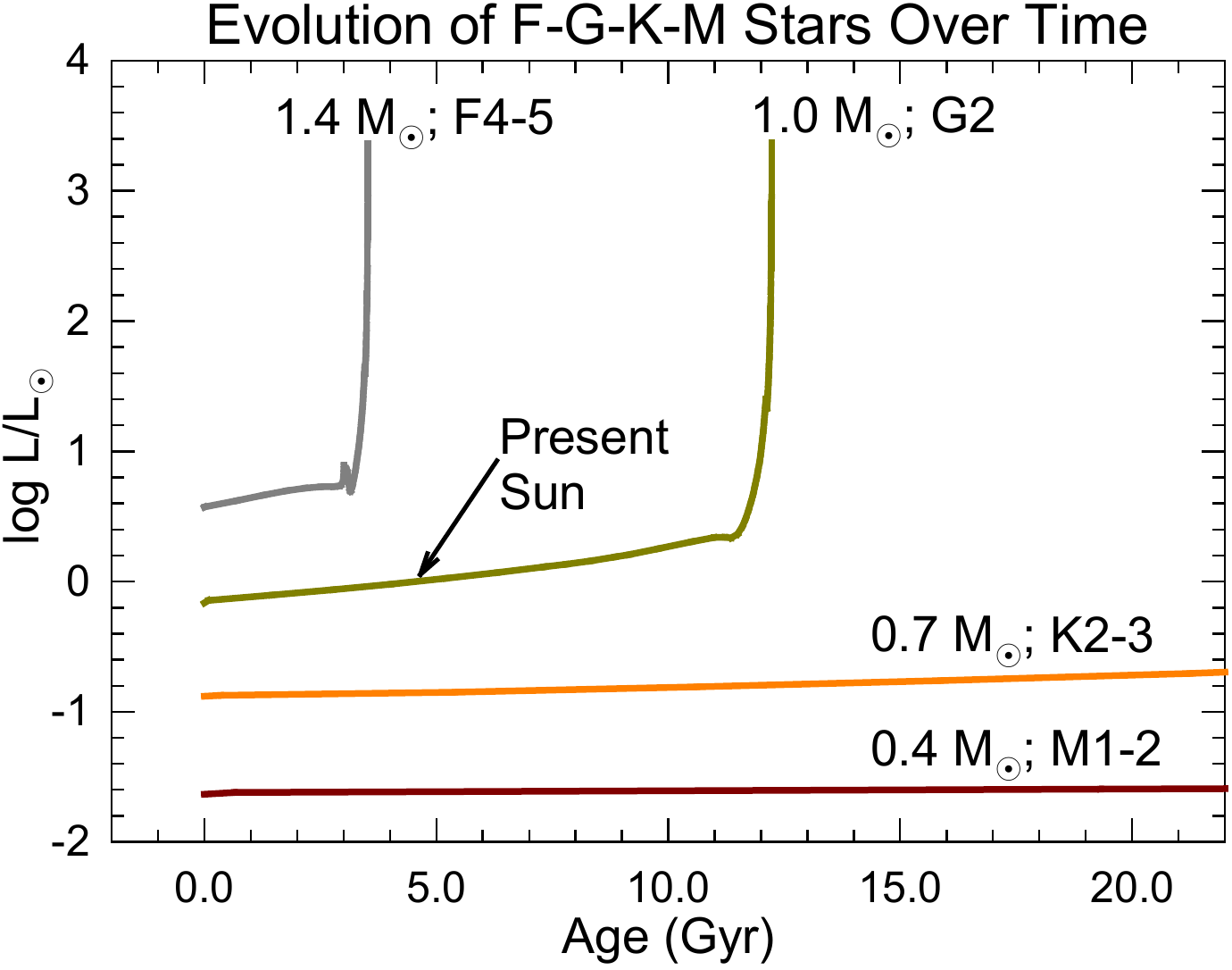}
\end{center}
\vspace{-6mm}
\caption{{\bf Left -- (a) --} Inventory of known stars within 10-pc,
binned by spectral type. Note: dM stars represent $>$75\% of the main-sequence stars.
{\bf Right -- (b) --} Luminosity changes over time for representative
F5~V, G2~V, K2~V and M1--2~V stars. The evolution tracks are from BaSTI
(http://albione.oa-teramo.inaf.it/).  Note: the luminosities of the
lower mass dK and dM stars change very slowly with time.}
\vspace{-5mm}
\end{figure}

\section{The {\em Living with a Red Dwarf} Program}
We have extended our ongoing {\em Sun in Time} program \citep{guinan2003,guinan2009}
 on solar dynamo physics, angular momentum loss,
Age-Rotation-Activity relations \& XUV (X-ray -- UV; $\sim$10--3200\AA) 
irradiances (and effects on planetary atmospheres and life) of solar type stars 
to more numerous, cooler red dwarfs.  
With the support of the U.S. National Science Foundation (NSF), 
we have been carrying out a comprehensive study of main sequence K/M stars
across the electromagnetic spectrum.  This {\em Living with a Red Dwarf} program includes
 $\sim$400+ nearby dK/M stars with a wide range of ages ($\sim$0.1--12 Gyr) 
and rotations ($P_{\rm rot}$ $\approx$ 0.4--190-d) that have vastly different 
levels of age/rotation dependent chromospheric UV, Ca {\sc ii} $HK$ and H$\alpha$ 
emissions, as well as corresponding transition region (TR) FUV--UV and coronal
X-ray emissions.  So far, from ground based photometry, we have determined rotation 
periods (from star spot brightness modulations), star spot fill-factors and flare frequencies 
for $\sim$140 of these stars \citep{engle2009}. Sample light curves are shown in Fig. 2. 

In addition, we are utilizing the HEASARC X-ray ({\em ROSAT}, {\em XMM} and {\em Chandra}) and
MAST FUV--UV ({\em IUE}, {\em HST} and {\em FUSE}) archives to determine the XUV fluxes for a subset of the program stars.
To fill in gaps in coronal X-ray coverage, we are currently observing a sample
of dM stars (with reliable ages) with the {\em Chandra X-ray Observatory}.  As in
the {\em Sun in Time} program, we are developing X-ray--UV irradiance measures for
red dwarf stars.  Our study (and others) shows that young red dwarf stars rotate rapidly 
and subsequently lose angular momentum over time, slowing their rotation and
consequently weakening their magnetic dynamos, resulting in significantly 
diminished coronal X-ray and chromospheric UV emissions with stellar age. There is also
evidence for a ``funneling effect'' in G/K/M stars, where stars $>$1 Gyr in age show a tighter
correlation between age and rotation as compared to stars $<$1 Gyr, which display a good bit of
scatter. The compensation mechanism is such: if you take two ZAMS K/M stars -- and Star A is rotating much
more rapidly than Star B -- then Star A will have a more robust dynamo with stronger winds, and will lose
angular momentum (and spin-down) more rapidly than Star B. This mechanism will bring the two stars
to a similar rotation rate as they age.

\begin{figure}
\begin{center}
\includegraphics[width=0.75\textwidth]{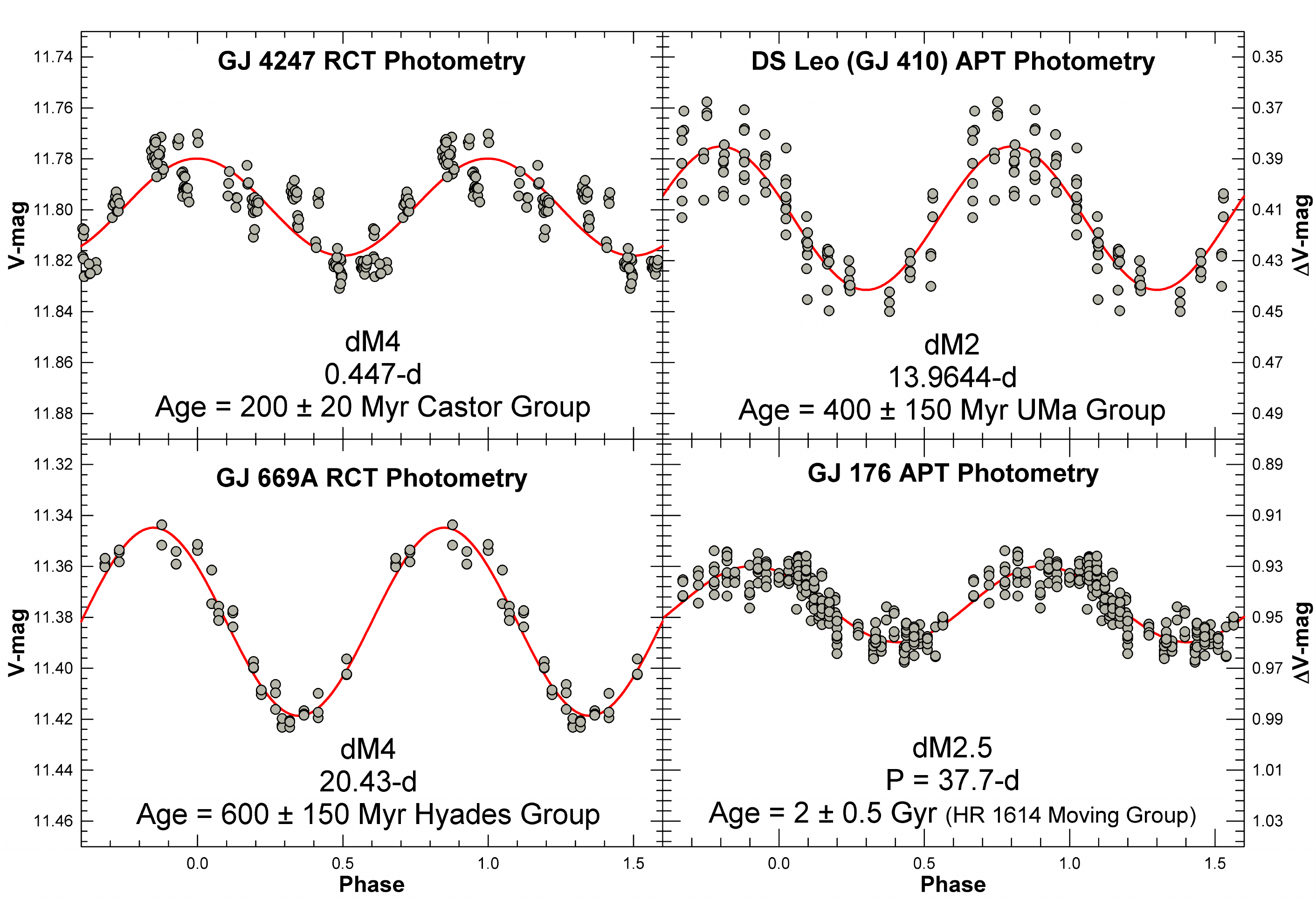}
\end{center}
\vspace{-6mm}
\caption{Representative V-band light curves are shown for four program dM2--4 stars. The
photometry was carried out using the 1.3-m \emph{RCT} \& 0.8-m \emph{FCAPT} telescopes. The stars have ages and
corresponding rotation periods of -- 200 Myr; P$_{\rm rot}$ = 0.447-days -- to -- 2.0 Gyr;
P$_{\rm rot}$ = 37.7-days.}
\vspace{-5mm}
\end{figure}

\section{Age-Rotation-Activity Relationships for dM Stars} 
\subsection{Photometrically Determined Rotation Periods for dM stars}
We have been carrying out photometry (or utilizing archival photometry) of a 
representative sample of $\sim$400 K/M stars to determine rotation periods,
starspot and flare properties. As found in this study, the rotation periods of K/M stars
are indicative of their ages (see Fig. 3a).  In fact, rotation rates are a more direct 
indicator of age than activity measures like Ca {\sc ii}, H$\alpha$ or $L_{\rm X}$, 
because these measures can be affected by: rotation effects (starspots/plages), 
activity cycles \& flaring. But the spin-down is dependent on spectral type (mass and 
convective zone (CZ) depth). However, in very young populations (ages $<$ 0.5 Gyr), the
rotation periods of red dwarfs typically range from $<$0.5-days up to $\sim$10 days (most
less than 5-days) for the same age bracket -- e.g. \citet{mess2010}. This arises from differing
initial conditions and the inclusion of pre-main sequence stars and close binaries in the
sample. Also in this study, we exclude spectral types later than $\sim$M7 because they appear to have 
Age-Rotation-Activity relations quite different to earlier spectral types. From a limited sample of 
M7--9 V stars, it appears that they do not undergo magnetic breaking over time.

Young K/M stars typically spin rapidly and have correspondingly robust dynamos \& magnetic
activity. Over time, though, they lose angular momentum via magnetized stellar winds and 
their rotation periods lengthen and their (magnetic-induced) activity decreases. This is 
illustrated in Fig. 3a showing the $P_{\rm rot}$-Age relation for program dM0--5 stars, and Fig. 3b shows the 
(coronal X-ray) $log~L_{\rm X}$-Age relation.  Photometry is being conducted
using the 0.8-m {\em Four College Automatic Photoelectric Telescope} ({\em FCAPT}) and the 1.3-m 
{\em Robotically Controlled Telescope} ({\em RCT}), both located in Arizona. This study has uncovered 
low amplitude rotational light modulations (and thus rotation periods) for many of these stars. 
There is also compelling  evidence of long-term light and activity variations in some of these stars (e.g. Proxima Cen -- $P_{\rm cyc}$ $\approx$ 7.1-yr -- from both light and X-ray variations) 
indicative of solar-like magnetic activity cycles.  We have also been searching for light variations in additional equatorial and southern  K/M stars included in the {\em All Sky Automated Survey} (\emph{ASAS-3}; \citet{pojmanski2001}). Utilizing the period search routines in the latest version of the \emph{Period Analysis Software} (\emph{Peranso} -- \url{http://www.peranso.com}), strong evidence of periodicity and long-term systematic variations in brightness have been uncovered for dozens of additional K/M stars,
including Proxima Cen. To broaden the database, we have recently been utilizing ultra-high
precision, time series photometry of $\sim$1000+ K/M stars from the \emph{Kepler} Mission to determine
precise rotation periods and unprecedented information on starspot genesis and evolution, as well as new information on flaring frequencies of red dwarfs. These results will be discussed in an upcoming report.

\subsection{Age by Association: Determining the Ages of K/M Stars}
It is crucial for this study that we have reliable age estimates to study the evolution 
of angular momentum, magnetic activity \& XUV spectral irradiance. However,
securing reliable ages of field K/M stars is next to impossible for red dwarf stars 
because of their extremely slow nuclear evolution. Once reaching the ZAMS, 
the stars' measureable properties -- such as $M_{\rm V}$, $T_{\rm eff}$, $log~g$ are essentially 
constant over time scales of the tens of billions of years (Fig. 1b). The essentially fixed 
luminosities of K/M stars (and thus fixed HZs) could be favorable for the formation and evolution 
of life on possible hosted HZ planets. But this is an obvious drawback for determining 
ages of red dwarfs from isochronal fits (in $L$--$T_{\rm eff}$ space). For example, 
over the $\sim$4.6 Gyr lifetime of the Sun, its luminosity has increased by $\sim$30\%. 
By contrast, a mid-dM star would undergo a luminosity increase of $<$1\% over the 
same time period. Until recently, the ages of K/M stars could only be reliably determined 
from memberships in nearby star clusters or moving groups with reliable ages 
(almost entirely $<$ 2 Gyr), or statistically inferred from sufficiently high $UVW$ space motions, 
indicating either Old Disk ($\sim$7--10 Gyr) or Halo ($\sim$10--13 Gyr) ages.  For example, 
ages for young (200--650 Myr) K/M stars can be found from memberships in nearby 
open clusters such as the Castor moving group ($\sim$200 Myr), the Ursa Major (UMa) 
moving group ($\sim$300--550 Myr), the Hyades cluster ($\sim$625 Myr) and the HR 1614 
moving  group ($\sim$2 Gyr). The Rotation-Age and $log~L_{\rm x}$-Age relations (for M0 V -- M5 V stars
only) are shown in Figure 3.  

A recent, important ``age method'' for red dwarfs is membership in wide binaries 
and common proper motion pairs where the companion star's age is known from
evolutionary tracks or from white dwarf (WD) companions (from stellar evolution 
and cooling times).  For example, \citet{sil2005} obtained ages of a number 
of red dwarfs with WD companions, but these ages suffered from large uncertainties
 since $V-I$ and $B-V$ colors were used to derive the WD cooling ages from models 
and, more importantly, a typical WD mass of $0.6M_\odot$ was assumed. Now, 
however, \citep{catalan2008,garces2011,zhao2011} made 
breakthroughs in the reliability of determining WD ages from cooling times and stellar 
evolution theory, utilizing more accurate, spectroscopically determined 
WD $T_{\rm eff}$ \& $log~g$ values which yield reliable masses. Combined with 
modern evolution and WD cooling theory (Salaris et al. 2000), the total WD age (lifetime 
of progenitor star + WD cooling time) is accounted for. An outline of the method is 
given in Fig. 4.  This new method has finally made it possible to determine ages for 
stars with no cluster membership or extreme space motions, allowing anchor points 
along our relationships where none could previously exist.  This is an important 
advance that significantly enhances the relationship reliability. The Age-Rotation-Activity
relations (Fig. 3) should improve as additional older stars are added. A thorough overview
of stellar age determinations is given in \citet{sod2010}.

\begin{figure}
\begin{center}
\includegraphics[width=0.48\textwidth]{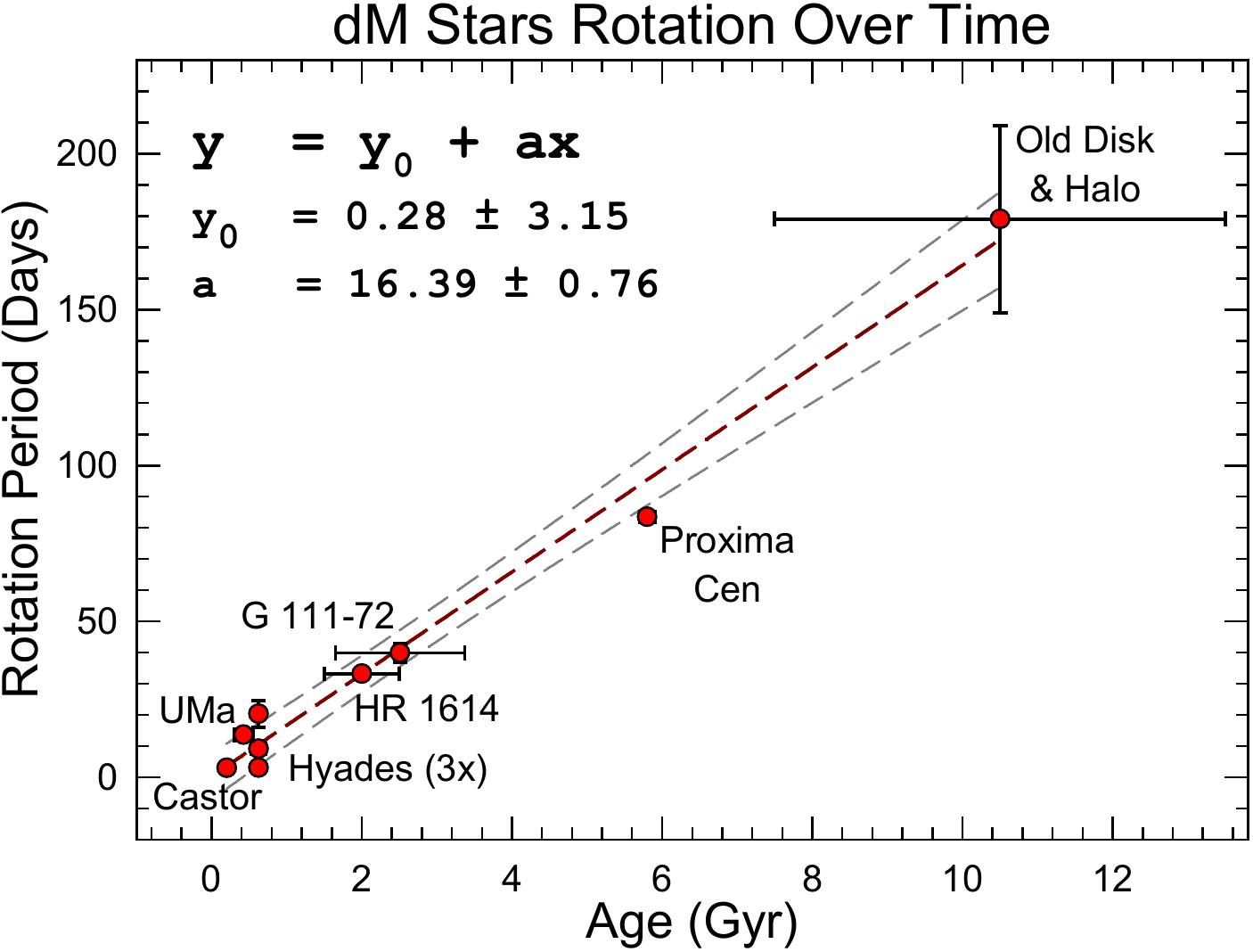}
\includegraphics[width=0.48\textwidth]{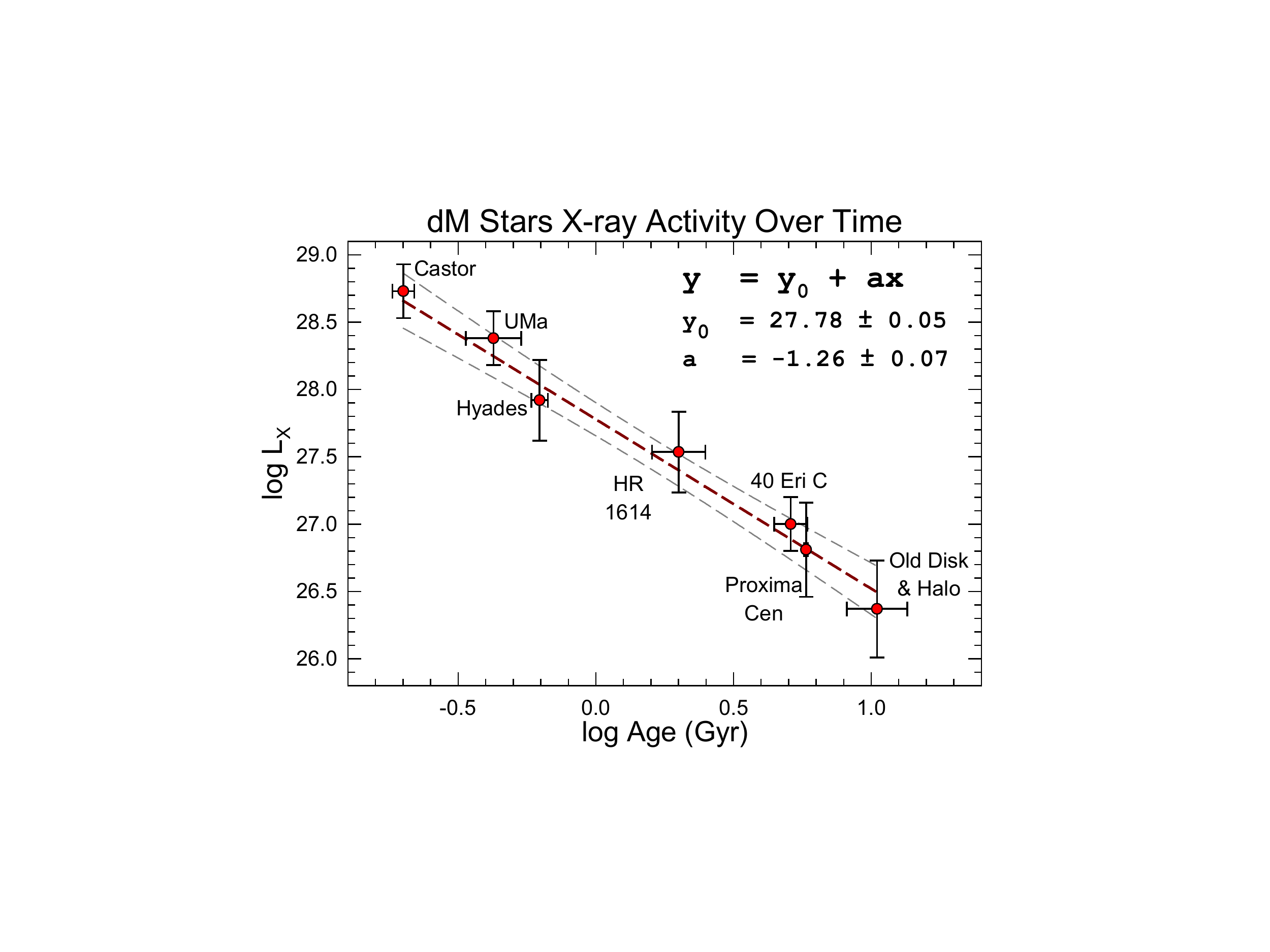}
\end{center}
\vspace{-6mm}
\caption{{\bf Left -- (a) --} The photometrically determined rotation rates
of dM stars are plotted vs. age. The data thus far indicates a linear trend,
in surprising contrast to what has been found for dG and dK stars.
{\bf Right -- (b) --} The decrease in coronal X-ray emission ($log~L_{\rm X}$) of dM stars
is shown. Saturation occurs at \emph{age} $<$ 0.3 Gyr}
\vspace{-5mm}
\end{figure}

\subsection{Examples of Age-Rotation-Activity Relation Calibration and Applications}

Here we present several example applications of our Age-Rotation-Activity
relations, and discuss the calibration of these relations with the 40 Eri ABC and G 111-72 
systems. Both of these star systems have WD components with reliable ages.  We also illustrate
the age determinations of two (the only two so far) exoplanet systems (GJ 581 and HD 85512)
that host super-Earths orbiting within the host K/M star's HZ, and discuss a new result for the
short-period transiting Hot Jupiter around HD 189733, whose true age and dynamic nature are
inferred from the age determination of a recently identified faint M4--5 V companion. We also
discuss the age determination for GJ 1214 -- a newly discovered transiting, short-period, hot
super-Earth planet system.

\noindent{\bf Applying WD evolution+cooling time ages to the 40 Eri \& G 111-72 systems:}

40 Eri ABC is a nearby (16.5 ly) triple star system that consists of a 4.4-mag K1 V star,
a 9.5-mag DA4 white dwarf and a 11.2-mag M4.5 V star.  The WD component, 40 Eri B, has
well-determined physical properties ($T_{\rm eff}$, $M/M_\odot$, $L/L_\odot$ and $log~g$)
that permit its age to be reliably calculated from the WD cooling time and progenitor
evolutionary age.  Two recent independent
age determinations for the 40 Eri B are essentially identical, yielding an age of
5.1$\pm$0.7 Gyr \citep{ballouz2010,zhao2011}. We used this WD age
to calibrate the measured Ca {\sc ii} $HK$ and coronal X-ray luminosity ($L_{\rm X}$) of
the K1 V and M4.5 V stars.  We have been carrying out photometry to determine the
photometric rotation periods for these benchmark stars.

We have also been carrying out photometry and spectroscopy of 
wide binary WD + K/M system that have good WD age estimates. So far this has resulted in rotation periods for the dM-star components of several WD+dM wide pairs with ages. For example, G 111-72 
($age$ $\approx$ 2.5 Gyr) has been found to have a photometric rotation period of 
$P_{\rm rot}$ =  39.86-days.  These stars are plotted in Fig. 3). Several of these stars also have been approved for X-ray observations by \emph{Chandra} 
during  2011/12.

\begin{figure}
\begin{center}
\includegraphics[width=0.95\textwidth]{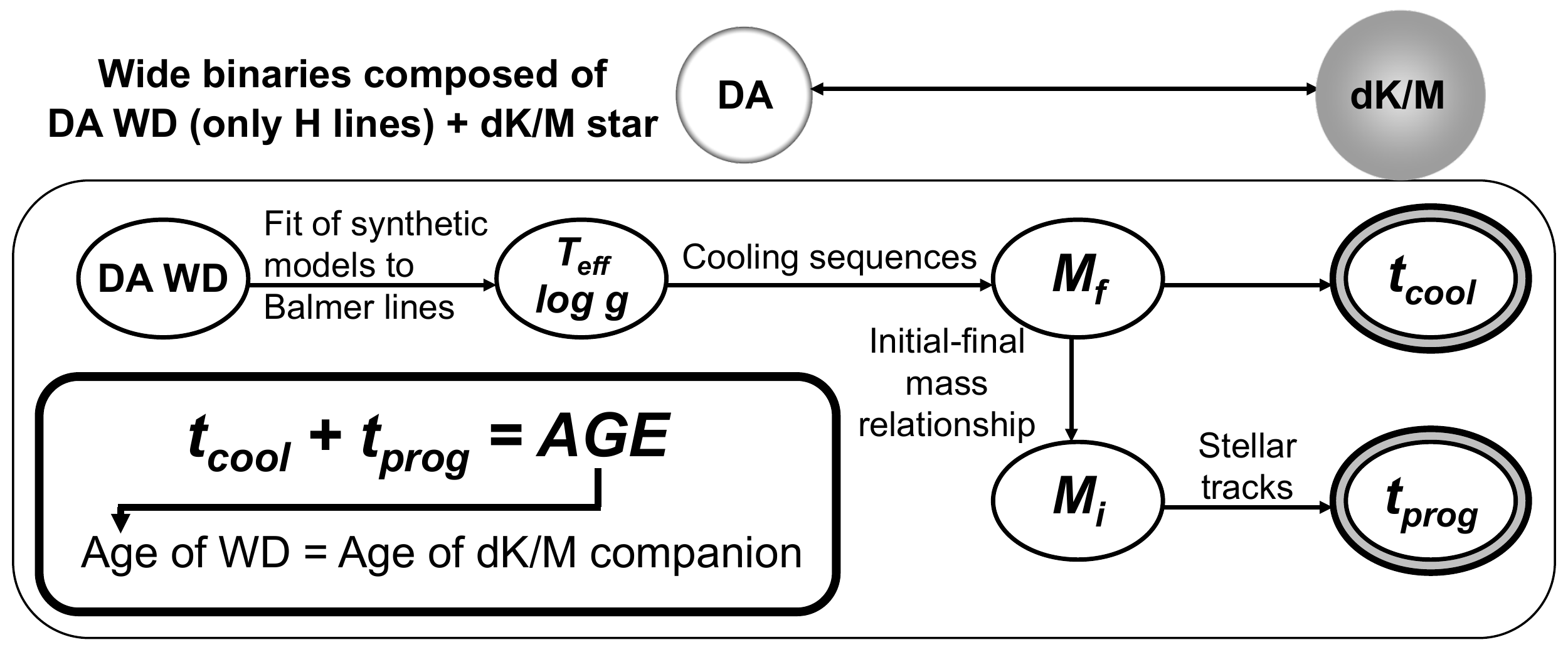}
\end{center}
\vspace{-6mm}
\caption{Method used to determine ages of DA WD + dK/M star binaries. Based on the work
of \citet{catalan2008}.}
\vspace{-5mm}
\end{figure}

\noindent{\bf Ages of super-Earth Planets Hosted by Red Dwarfs -- Examples:} 

\noindent {\bf GJ 581:} The nearby M3 V red dwarf GJ 581 (distance $\approx$ 20 ly) is one of the 
most famous host stars of potentially habitable planets. At present, GJ 581 is 
the nearest star system with super-Earth planets that could be suitable for life.  
The exact number of super-Earth type planets hosted by GJ 581 is still debated --
the most recent HARPS radial velocity study indicates the presence of four exoplanets \citep{for2011}
instead of the six planets previously reported by \citet{vogt2010}.
Noteworthy, two of the large Earth-mass planets, GJ 581c ($P$ = 12.92-d; $a$ = 0.07 AU; $M~sin~i$ = 5.32
$M_\oplus$) and GJ 581d ($P$ = 66.6-d; $a$ = 0.22 AU; $M~sin~i$ = 6.1 $M_\oplus$) remain of great
interest because they are located close to the inner (hot) edge and outer (cold) edge of the M3 V
star's habitable zone (Fig. 5a). Depending on the atmospheric properties of the planets
(atmospheric composition/circulation, density and albedos) these planets could have conditions favorable for the
development of life (Fig. 5b). Habitability also depends on the age and activity of the
host star.  From our rotation-age relation, we estimate an $age$ = 5.7$\pm$0.8 Gyr from the
$P_{\rm rot}$ = 93.2$\pm$1.0 day reported by \citet{vogt2010}. This age estimate is in excellent
agreement with that of 4--8 Gyr indicated from the star's $L_{\rm X}$ upper limit of $log~L_{\rm X}$
$<$ 26.4 ergs/s and Ca {\sc ii} $HK$ emissions. 

\noindent {\bf HD 85512:}  Recently, a large Earth-size planet has been found orbiting the
nearby K5 V star HD 85512 \citep{pepe2011}. This is important since, like GJ 581c, HD 85512b
orbits near the inner edge of the star's HZ and could have conditions suitable for life \citep{kalt2011}.
HD 85512 has a slow rotation period (for mid-dK stars) of 47.1$\pm$7 days \citep{pepe2011}. This period,
if correct, indicates an age of 8$\pm$2.5 Gyr (from our $P_{\rm rot}$-Age relation for early/mid-dK
stars \citep{wolfe2010}; somewhat older than
the Ca {\sc ii} $HK$ ($R'_{HK}$) age estimate of 5.6$\pm$0.6 Gyr by \citet{pepe2011}).

\noindent {\bf GJ 1214:} Another system to which our method has been applied is that of 
GJ 1214 -- a dM4 star with a transiting, hot super-Earth. The planet has a mass of 
6.45 M$_\oplus$, an orbital period of 1.58-days and an estimated equilibrium temperature 
of $\sim$500 K \citep{berta2011}. Utilizing photometry from the MEarth Program, 
\citeauthor{berta2011} found a likely stellar rotation period of $\sim$53-days. Applying our 
$P_{\rm rot}$--Age relation,  we find $age$ = 3.2$\pm$0.6 Gyr, which agrees with the $>$3 Gyr lower age 
limit of \citet{berta2011}.
 
\noindent{\bf Evidence of Host Star Spin-Up by Close-In Hot Jupiters: the Case of HD 189733:} Our
recent study of the bright, short period transiting exoplanet system HD 189733 provides an excellent
example of what new information can be found once the age of the star is determined \citep{san2011}.
HD 189733 A is a K2 V star that has attracted much attention because it hosts a
transiting, hot Jupiter-exoplanet. HD189733b has one of the shortest known orbital-periods 
($P$ = 2.22-days) and is only 0.031 AU from its host star \citep{bouchy2005}. HD 189733 A has a $P_{\rm rot}$ $\approx$ 12-d, coronal L$_{\rm x}$ $\approx$
10$^{28}$ ergs/s, and moderate-to-strong chromospheric Ca {\sc ii}-HK emission, all indicating an $age$
$\approx$ 0.6--1.0 Gyr (as inferred from our Age-Rotation-Activity relations for early dK stars.
However, this age is discrepant with an older-age inferred from the star's low Lithium-abundance
($\frac{1}{10}$ Solar). But the Age-Rotation-Activity determination assumes no tidal effects from
companions -- such as a close planet. \citet{bakos2006} discovered a dM4 companion star (HD
189733 B: 12'' distance to the K2 V star). $XMM-Newton$ observations of HD 189733 A\&B
carried out recently by \citet{pill2010}, surprisingly revealed that HD 189733 B shows no
X-ray emission, with an upper limit of $L_{\rm X}$ $<$ $9x10^{26}$ ergs/s. Our $L_{\rm X}$-Age
relationship (Fig. 3b) indicates an age for the dM4 star of $>$ 4 Gyr.
Also the lack of H$\alpha$ emission and weak Ca {\sc ii} $HK$ emission in the dM star's spectra indicate
weak magnetic activity consistent with slow rotation and a lower limit age of $\sim$4 Gyr.
These age proxies indicate that the binary is indeed old and that the fast
rotation and corresponding high levels of activity for the K2 V star can be resolved if this star
has been spun-up by its nearby planetary companion. Thus, the planet's orbital angular momentum has been transfered to
HD 189733 A via tidal and/or magnetic interactions. \citet{cohen2011} have recently discussed evidence for
magnetic interactions between the planet and the host star as well as the probable effects of tidal
and magnetic interactions. This loss of orbital angular momentum by the planet should eventually
cause the planet to spiral inward and be disrupted near its Roche lobe.

\begin{figure}
\begin{center}
\includegraphics[width=0.48\textwidth]{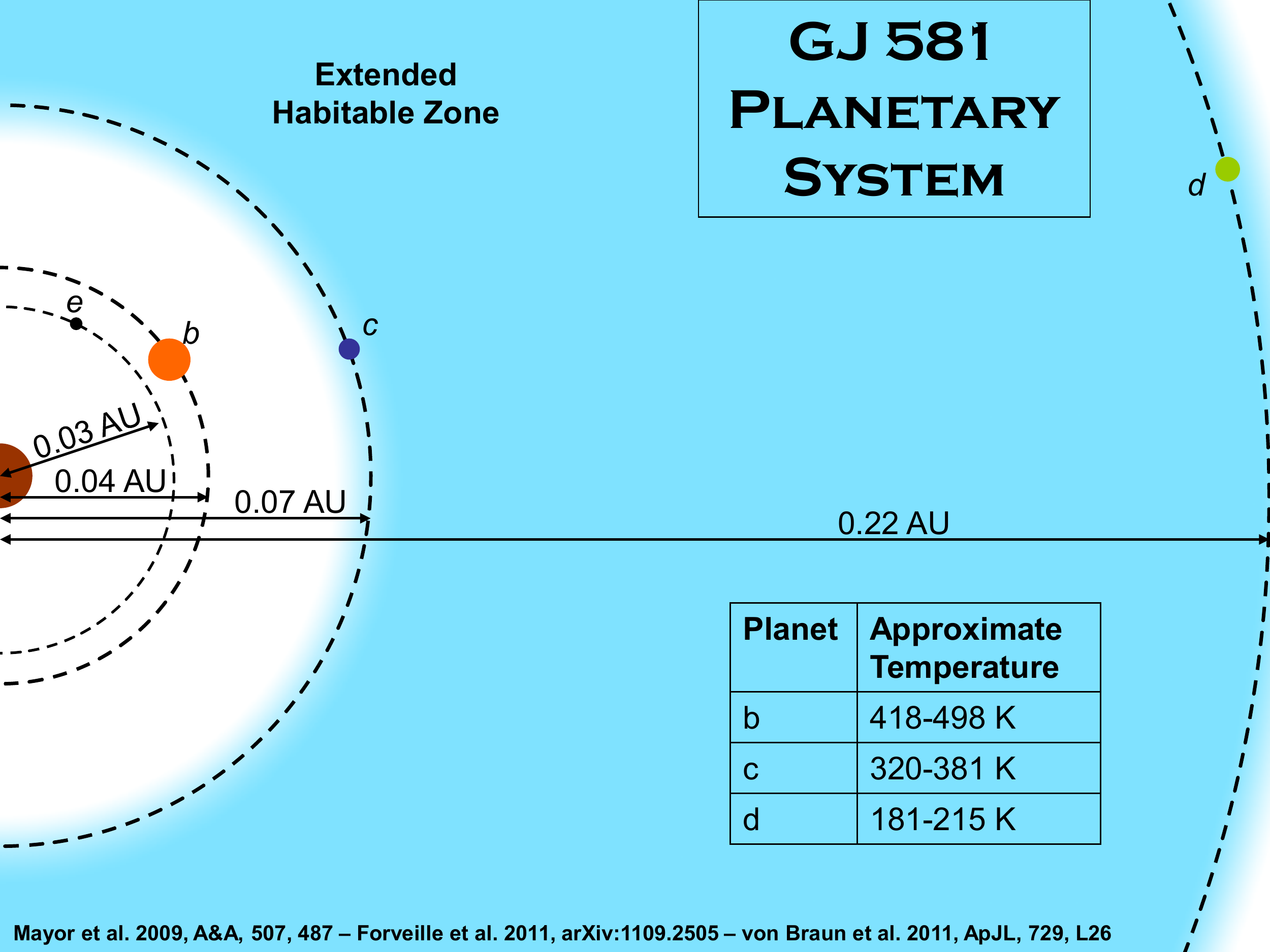}
\includegraphics[width=0.48\textwidth]{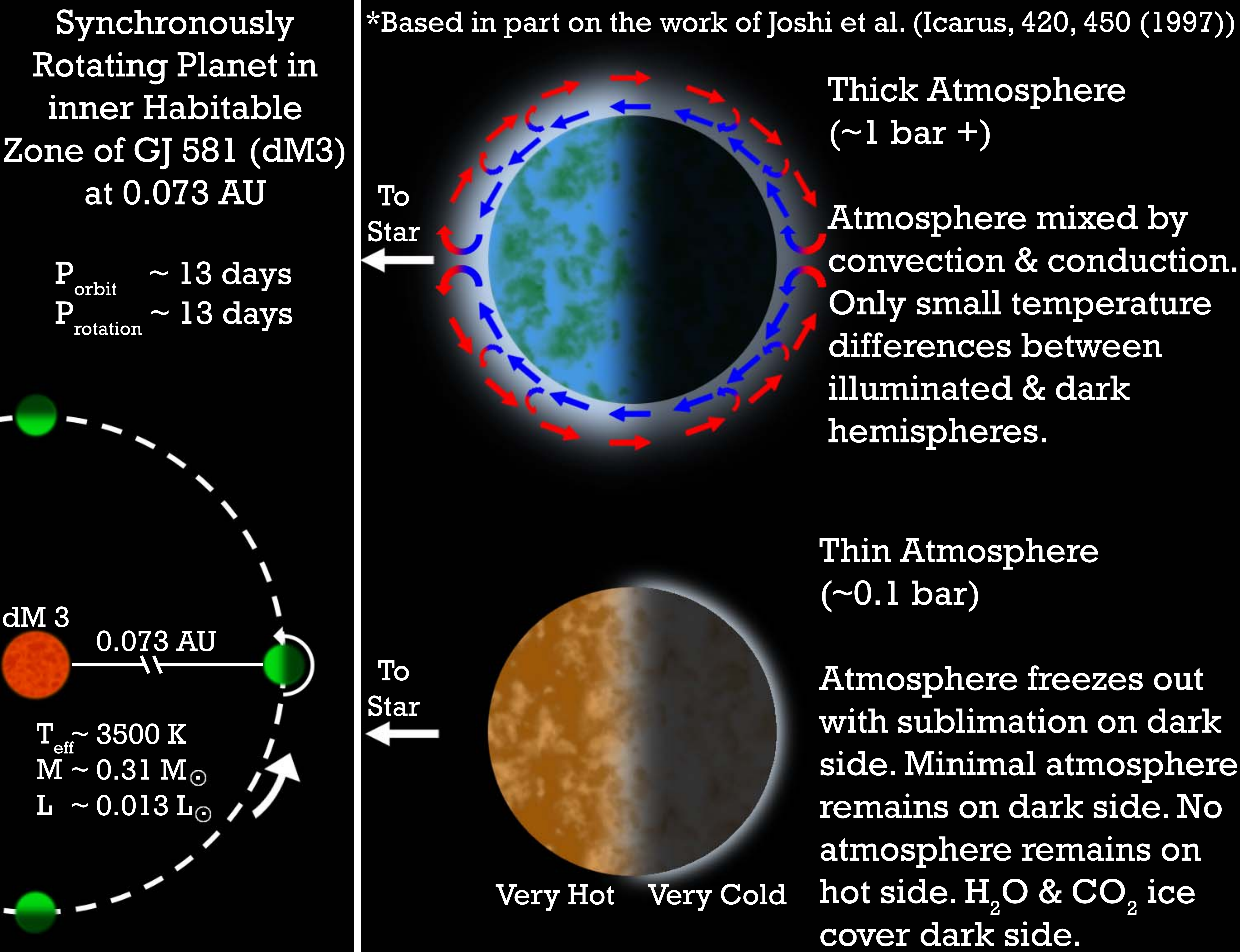}
\end{center}
\vspace{-6mm}
\caption{{\bf Left -- (a) --} Illustration of the GJ 581 planetary system, plotting the
orbits of GJ 581e, b, c and d.  Super-Earths ``c'' and ``d'' are located near the inner
(warm) and outer (cold) edges of the star's habitable zone.
{\bf Right -- (b) --} Illustration of planet GJ 581c showing the synchronization
of its $\sim$13-d orbital and rotation periods. The thermal effects of a thin ($<$0.1 Bar)
and thick (1 Bar) atmosphere are shown. In the thick atmosphere case,
the dark hemisphere of GJ 581c could have temperatures suitable to support life
via atmospheric circulation. With a strong greenhouse effect, planet GJ 581d could
also be suitable for life on its starlit side.}
\vspace{-5mm}
\end{figure}

\subsection{Constructing K/M Star X-ray--UV (XUV) Irradiances}
A detailed characterization of K/M star radiation, plasma environments and changes
with time is very important to the study of extrasolar planets -- especially for Earth-size 
HZ planets with potential for life.  Because of their slow
nuclear evolution rates, the luminosity and $T_{\rm eff}$ of the stars and resulting
photospheric spectral radiances (from $\sim$3000\AA~out to IR wavelengths) remain
essentially fixed with time for a given mass. However, the radiation at shorter
wavelengths (the XUV region) arises from dynamo-generated coronal, TR and
chromospheric emissions, and undergoes a measurable decrease as the stars lose angular momemtum
with age (Fig.6a). XUV radiation can profoundly affect planetary atmospheres, due to 
photoionization and photochemical reactions, along with ion pickup processes from stellar winds impacting 
planetary thermospheres \citep{lammer2003,grie2004,ribas2005,tian2009}. 

The XUV irradiance tables ($\sim$10--3200\AA) we are developing will be important 
for studying the evolution of the atmospheres of planets hosted by K/M stars. The 
FUV/NUV irradiances of representative G, K, and M stars are shown in Fig. 6b. In particular, the age 
sequence of spectral irradiances from our sample (e.g. Fig. 3b for coronal X-rays and 
Fig. 6a for FUV variations with age) can delineate not only the present state of the new 
extrasolar planets detected, but also the evolution of their atmospheres 
and can assess the possibility of life. One major effect that 
XUV radiation can have (in combination with stellar wind interactions) is the erosion 
of the hosted planet's atmosphere \citep{lammer2003,grie2004}.  
As discussed by \citet{grie2004}, close-in planets without strong 
(protective) magnetic fields are especially susceptible to atmospheric
erosion \& loss by the star's XUV and wind (plasma) fluxes.  Also, the frequent
flaring of dM stars (especially those with ages $<$ 1 Gyr) and tidal locking of
close-in planets could challenge the development of life. If life were to form,, however, the long
lifetimes of the host K/M star could be favorable to the development of
complex (possibly even intelligent) life. Numerous dM stars in the solar neighborhood
are old ($>5$ Gyr), presenting possibilities for highly advanced forms of intelligent 
life \citep{tarter2007}. Additionally, young red dwarf stars undergo frequent flares, in which
their UV fluxes increase 10--100$\times$ for several minutes. The increased XUV 
flare radiation could have further adverse effects on the retention of a planet's atmosphere 
and be harmful to possible life on its surface. Recently, however, \citet{tian2009} evaluated the
ability of a 6--10 $M_\oplus$ super-Earth to retain a primary CO$_2$ atmosphere while orbiting
a dM star. \citeauthor{tian2009} found that the atmosphere could be retained, even for dM star
XUV activity up to 1000$\times$ that of the Sun.

\begin{figure}
\begin{center}
\includegraphics[width=0.48\textwidth]{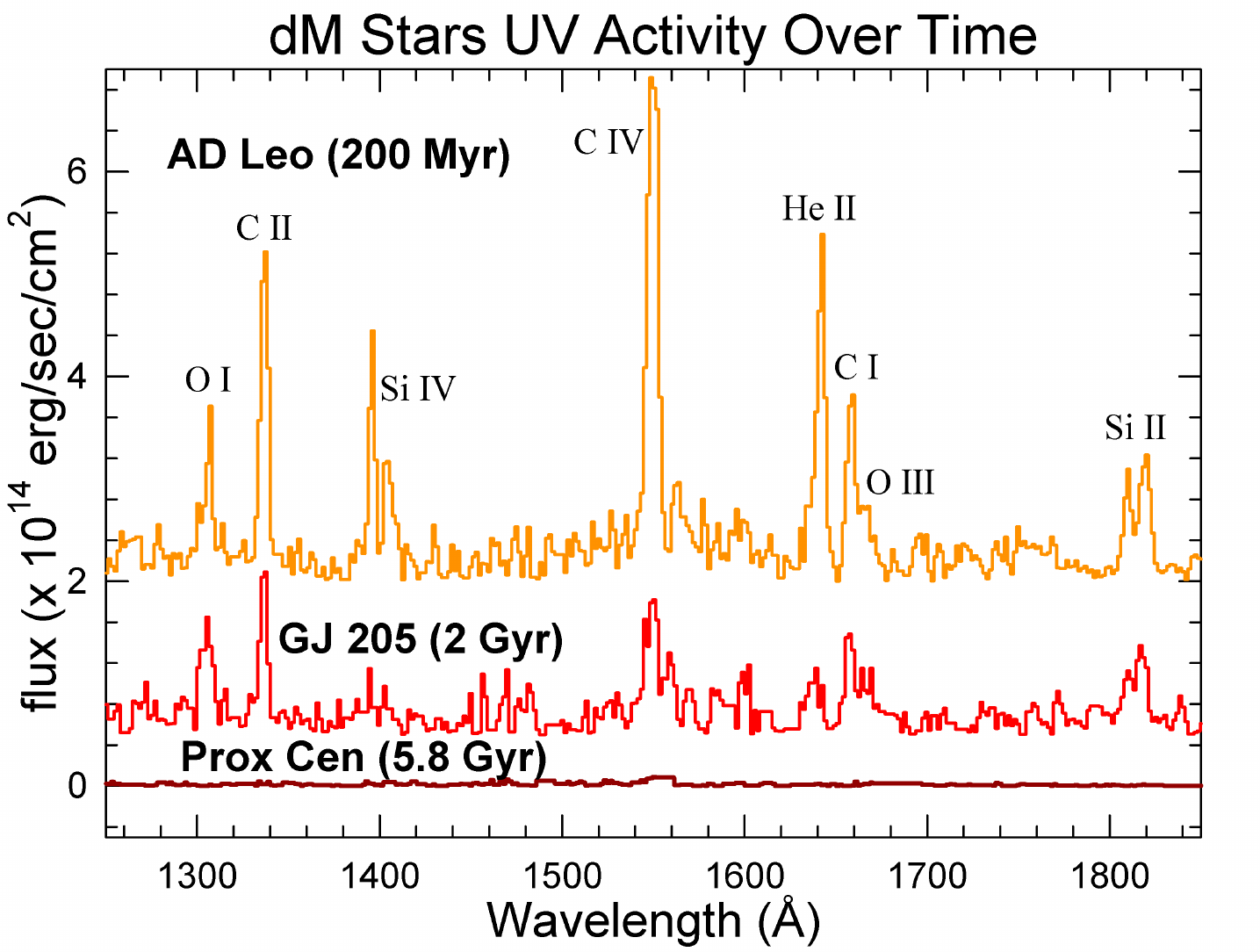}
\includegraphics[width=0.49\textwidth]{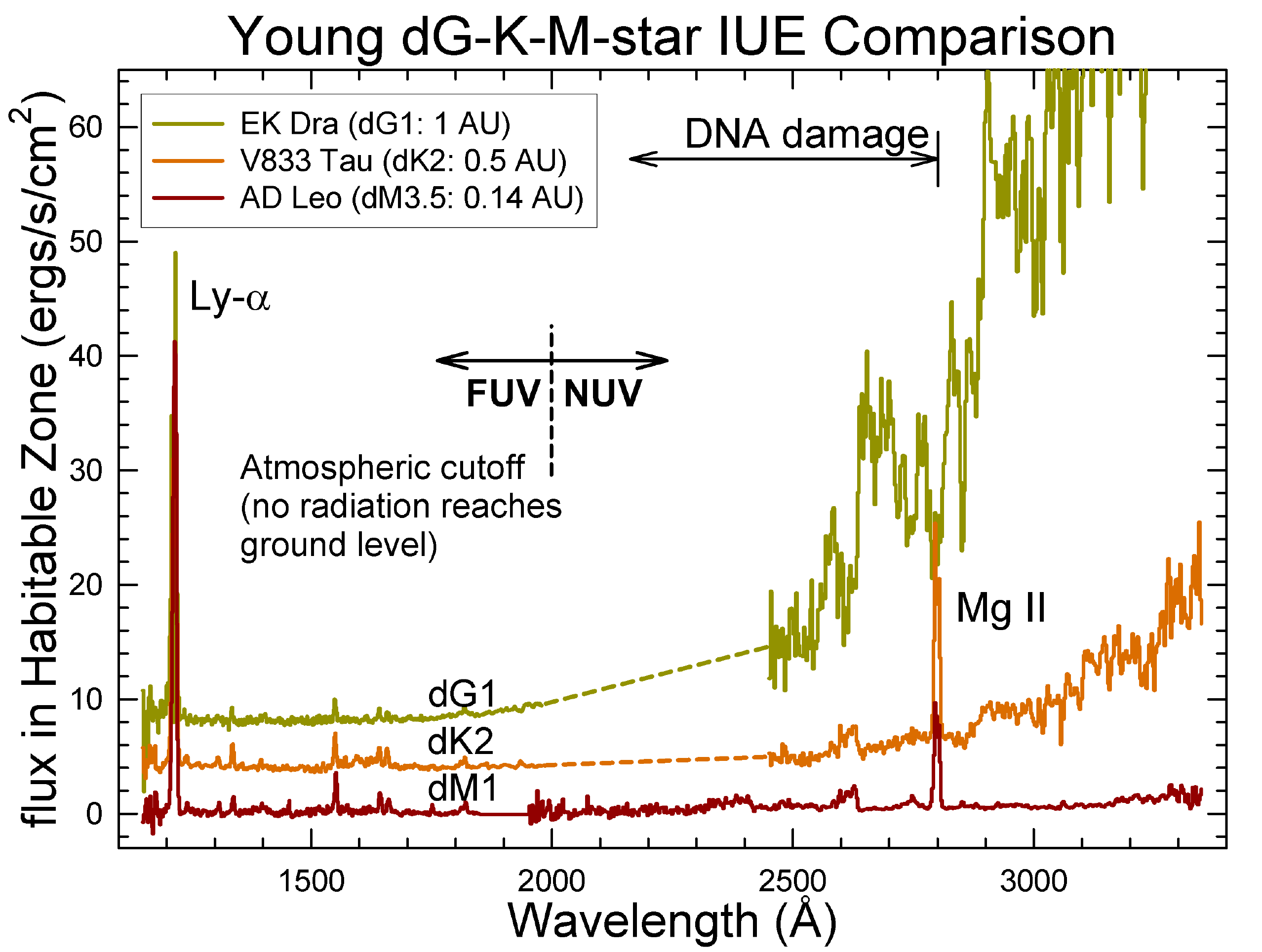}
\end{center}
\vspace{-6mm}
\caption{{\bf Left -- (a) --} The FUV (1250--1850\AA) spectrophotometric
fluxes (adjusted for distance) of three dM stars of different
ages are shown. The chromospheric and TR line emissions decrease significantly 
(down to $\sim\frac{1}{100}$ the original level) over an age range of 0.2--5.8 Gyr. 
{\bf Right -- (b) --} The FUV (1200--2000\AA) and NUV (2000--3200\AA) irradiance plots
computed for the respective habitable zones for young G, K and M stars are
shown. In the FUV, line emissions dominate and are nearly equal 
for these stars. But in the NUV, photospheric continuum flux dominates
for the G1 V and K2 V stars; for the red dwarf there is essentially no NUV radiation
outside of flares.}
\vspace{-5mm}
\end{figure}

\section{Conclusions \& Future Prospects}
The {\em Living with a Red Dwarf} program, when completed, should provide valuable data
for the study of exoplanets by providing the age estimates, magnetic evolution and
resulting XUV emissions of their host stars. The Age-Rotation-Activity relationships 
for red dwarf stars found through this program will allow galactic red dwarfs to have
reliably determined ages based on photometric rotation periods or measures of
X-ray, Ca {\sc ii} $HK$ ($R'_{\rm HK}$) or H$\alpha$ emissions (and accurate spectral
type for best precision). This could be especially important in realizing ages (from
rotation) for the thousands of red dwarfs being observed by \emph{Kepler}. In addition, as
part of the {\em Sloan Digital Sky Survey} program, about 50,000 spectra of K/M stars have been secured
\citep{boch2011}. We are also developing H$\alpha$-Age emission relations for program
stars with Stella Kafka (Carnegie Inst., DTM). From results in hand,
there are good (but preliminary) correlations between H$\alpha$ emission and age for
subsets of dM star, in which H$\alpha$ emission decreases with
age. Surprisingly, our program K/M stars, although smaller with deeper CZs and possibly even different
dynamo mechanisms, behave/evolve in a similar manner (but achieving much longer rotation periods)
to solar-type stars. When this study is realized, it may be possible to estimate approximate ages of
the numerous \emph{SDSS} red dwarfs that have spectra -- an important tool for galactic structure studies. 
With the ages of red dwarf star planetary systems reliably known, the dynamic evolutionary 
history and stability of the system can then be assessed. Furthermore, the past, 
present and future XUV radiative environments that the planets will face can be 
estimated. Also under study is the characterization of frequent flares that red dwarf stars 
display, in which their UV fluxes increase 10--100$\times$ for several minutes. The increased XUV 
radiation from flares could have adverse effects on the retention of a planet's atmosphere
and be harmful to possible life on its surface. Finally, planetary habitability can be 
assessed based on the irradiance data and statistical possibilities can be assigned to
whether or not life could originate and evolve on a planet hosted by a red dwarf. 

Thus, potentially habitable planets around red dwarfs are clearly an important issue for future 
study. Dedicated search programs for planets orbiting K/M stars are being developed. For
example, the Carnegie \emph{Planet Finder Spectrograph} (\emph{PFS} -- $<$ 1 m/s precision
optical echelle) is now in use on the 6.5-m Magellan II telescope, and is
emphasizing K/M stars as its primary targets. Also, the \emph{High Accuracy Radial
velocity Planet Searcher} (\emph{HARPS}) instrument at La Silla achieves a similar precision,
has been operating since 2003 and has made numerous discoveries concerning both
planets and various aspects of stellar activity. Another ground-based
program is the the \emph{MEarth} project - a transit survey of $\sim$2000 dM stars
in the northern hemisphere \citep{irwin2009}. \emph{MEarth} can detect transits by bodies as
small as $\sim$2 $R_\oplus$.  Also, space-based planet search missions,
including \emph{CoRoT}, \emph{Kepler} and, in the future, \emph{Plato} and (maybe) \emph{Darwin/TPF}, are planning to target 
many additional red dwarf stars. It now appears timely to assess the likelihood of life not 
only around main sequence dG \& dK stars, like our Sun, but also around the very 
numerous, low luminosity dM stars.

\acknowledgements
This research is supported by NSF/RUI Grant No. AST 1009903 and NASA/Chandra 
Grant GO1-12024X which we gratefully acknowledge. Also, we thank the meeting organizers, in particular
Shengbang Qian, for a well-run, informative meeting in the beautiful and historic city of Lijiang, China.

%\bibliography{aspauthor}

\begin{thebibliography}{}

\bibitem[Bakos et al.(2006)]{bakos2006} Bakos, G.~{\'A}., 
P{\'a}l, A., Latham, D.~W., Noyes, R.~W., 
\& Stefanik, R.~P.\ 2006, \apjl, 641, L57 

\bibitem[Ballouz et al.(2010)]{ballouz2010} Ballouz, R.-L., Guinan, 
E.~F., Wasatonic, R., 
\& Engle, S.~G.\ 2010, Bulletin of the American Astronomical Society, 42, \#425.08 

\bibitem[Berta et al.(2011)]{berta2011} Berta, Z.~K., 
Charbonneau, D., Bean, J., et al.\ 2011, \apj, 736, 12 

\bibitem[Bochanski et al.(2011)]{boch2011} Bochanski, J.~J., 
Hawley, S.~L., \& West, A.~A.\ 2011, \aj, 141, 98 

\bibitem[Bouchy et 
al.(2005)]{bouchy2005} Bouchy, F., Udry, S., Mayor, M., et al.\ 2005, \aap, 444, L15 

\bibitem[Catal{\'a}n et 
al.(2008)]{catalan2008} Catal{\'a}n, S., Isern, J., Garc{\'{\i}}a-Berro, E., et al.\ 2008, \aap, 477, 213 

\bibitem[Cohen et al.(2011)]{cohen2011} Cohen, O., Kashyap, 
V.~L., Drake, J.~J., et al.\ 2011, \apj, 733, 67 

\bibitem[Engle et al.(2009)]{engle2009} Engle, S.~G., Guinan, 
E.~F., 
\& Mizusawa, T.\ 2009, American Institute of Physics Conference Series, 1135, 221 

\bibitem[Forveille et al.(2011)]{for2011} Forveille, T., 
Bonfils, X., Delfosse, X., et al.\ 2011, arXiv:1109.2505 

\bibitem[Garc{\'e}s et 
al.(2011)]{garces2011} Garc{\'e}s, A., Catal{\'a}n, S., \& Ribas, I.\ 2011, \aap, 531, A7 

\bibitem[Grie{\ss}meier et 
al.(2004)]{grie2004} Grie{\ss}meier, J.-M., Stadelmann, A., Penz, T., et al.\ 2004, \aap, 425, 753 

\bibitem[Guinan 
\& Engle(2009)]{guinan2009} Guinan, E.~F., \& Engle, S.~G.\ 2009, IAU Symposium, 258, 395 

\bibitem[Guinan et al.(2003)]{guinan2003} Guinan, E.~F., Ribas, 
I., \& Harper, G.~M.\ 2003, \apj, 594, 561 

\bibitem[Irwin et al.(2009)]{irwin2009} Irwin, J., Charbonneau, 
D., Nutzman, P., 
\& Falco, E.\ 2009, American Institute of Physics Conference Series, 1094, 445 

\bibitem[Kaltenegger et al.(2011)]{kalt2011} Kaltenegger, L., 
Udry, S., \& Pepe, F.\ 2011, arXiv:1108.3561 

\bibitem[Lammer et al.(2003)]{lammer2003} Lammer, H., Selsis, F., 
Ribas, I., et al.\ 2003, \apjl, 598, L121 

\bibitem[Mayor et 
al.(2009)]{mayor2009} Mayor, M., Bonfils, X., Forveille, T., et al.\ 2009, \aap, 507, 487 

\bibitem[Messina et 
al.(2010)]{mess2010} Messina, S., Desidera, S., Turatto, M., Lanzafame, A.C., \& Guinan, E.F.\ 2010, \aap, 520, A15 

\bibitem[Pepe et al.(2011)]{pepe2011} Pepe, F., Lovis, C., 
S{\'e}gransan, D., et al.\ 2011, arXiv:1108.3447 

\bibitem[Pillitteri et al.(2010)]{pill2010} Pillitteri, I., 
Wolk, S.~J., Cohen, O., et al.\ 2010, \apj, 722, 1216 

\bibitem[Pojmanski(2001)]{pojmanski2001}{Pojma\'{n}ski}, G. 2001, $ASPC$,
246, 53

\bibitem[Ribas et al.(2005)]{ribas2005} Ribas, I., Guinan, E.~F., 
G{\"u}del, M., \& Audard, M.\ 2005, \apj, 622, 680 

\bibitem[Santapaga et al.(2011)]{san2011} Santapaga, T., 
Guinan, E.~F., Ballouz, R., Engle, S.~G., 
\& Dewarf, L.\ 2011, Bulletin of the American Astronomical Society, 43, \#343.12 

\bibitem[Silvestri et al.(2005)]{sil2005} Silvestri, N.~M., 
Hawley, S.~L., \& Oswalt, T.~D.\ 2005, \aj, 129, 2428 

\bibitem[Soderblom(2010)]{sod2010} Soderblom, D.~R.\ 2010, \araa, 48, 581 

\bibitem[Tarter et al.(2007)]{tarter2007} Tarter, J.~C., Backus, 
P.~R., Mancinelli, R.~L., et al.\ 2007, Astrobiology, 7, 30 

\bibitem[Tian(2009)]{tian2009} Tian, F.\ 2009, \apj, 703, 905 

\bibitem[Vogt et al.(2010)]{vogt2010} Vogt, S.~S., Butler, 
R.~P., Rivera, E.~J., et al.\ 2010, \apj, 723, 954 

\bibitem[Wolfe et al.(2010)]{wolfe2010} Wolfe, A., Guinan, E.~F., 
Datin, K.~M., DeWarf, L.~E., 
\& Engle, S.~G.\ 2010, Bulletin of the American Astronomical Society, 42, \#423.10 

\bibitem[Zhao et al.(2011)]{zhao2011} Zhao, J.~K., Oswalt, 
T.~D., Rudkin, M., Zhao, G., \& Chen, Y.~Q.\ 2011, \aj, 141, 107 

\end{thebibliography}

\end{document}